\documentstyle[preprint,aps]{revtex}
\begin{document}
\draft
\title{Exchange interaction effects in inter-Landau level
 Auger scattering in a two-dimensional electron gas}
\draft
\author{E. Tsitsishvili$\dag$, Y. Levinson}
\address{Department of Condensed Matter Physics, The Weizmann Institute of
 Science,\\
 76100 Rehovot, Israel}
\draft
\maketitle
\begin{abstract}
We consider the influence of spin effects on the inter-Landau level
electron-electron scattering rate in a two-dimensional electron gas.
Due to the exchange spin splitting, the Landau levels are not  
equidistant. This leads to the 
suppresion of Auger processes and a nonlinear dependence
of the lifetime on the concentration of the excited
electrons even at very low excitation levels.
\end{abstract}
\hspace*{2cm} 
$\dag$Institute of Cybernetics, Euli 5, 380086 Tbilisi,Georgia\\
\pacs{72. 10.Di Scattering by phonons, magnons, and other nonlocalized
excitations
       73. 40.Hm Quantum Hall effect}
   
In the past years a considerable amount of work has been done on the
determination of the electron lifetime in the excited Landau levels (LL) in
a two-dimensional electron gas (2DEG)\cite{Sa,H,L,Ma,Hs}. It was found that
electron-electron $(ee)$ scattering is the dominant relaxation mechanism, when
the emission of LO phonons is suppressed off the magnetophonon resonance
conditions\cite{V,Mu}. In this case the electron lifetime is determined by
Auger processes, in which two excited electrons in the same LL are scattered,
 deexciting one to a lower LL, and exciting the second to a higher LL. The
decrease of the measured lifetimes with an increase in the $\it{excited}$ 
electrons' concentration, $n_{exc}$, has proved a convincing argument for
this conclusion.

One can think that the probability for an Auger process to occur,
 $\tau_{ee}^{-1}$, increases linearly with $n_{exc}$. However, the
experiments do not confirm this conclusion\cite{Ma}. We will show that
this "naive" picture is not complete, and that the nonlinear dependence
of  $\tau_{ee}^{-1}$ on $n_{exc}$ is due to spin effects.

In a 2DEG the LLs are equally spaced if one neglects spin effects (and
nonparabolicity). The exchange interaction violates the equidistant LL
 spacing. The exchange energy in the N$\sigma\:$ LL ($\sigma = \uparrow, 
\downarrow$) is usually presented as $- \Sigma_{{\rm N}\sigma}$, where
$\Sigma_{{\rm N}\sigma} =  E_{0}\, \nu_{{\rm N}\sigma}$, with $E_{0} > 0$,
 and $\nu_{{\rm N}\sigma}$ is the corresponding filling factor\cite{An}.
 The observed values of $E_{0}$ are of the order of a several meV, e.g.,
in GaAs $E_{0}$ = 3 meV $\div$ 6 meV at 10 T \cite{Us,vK,Ni,Dp}.
The  Zeeman spin splitting is important for the empty LLs only, since it
is much smaller ($\sim 0.2$ meV at 10 T) than the exchange energy. 
 
We will consider a situation that is similar to the experiment\cite{Ma}, in
which the 2DEG is spin polarized. At equilibrium the electrons occupy
the lowest LL, 0$\uparrow$,  with a filling factor $\nu_{0\uparrow} =
\nu <1$. Due to cyclotron absorption some of the electrons are excited
to the higher LLs, N$\uparrow$. Since the Auger processes preserve the 
initial spin orientation due to total spin conservation during the scattering
event, the LLs  N$\downarrow$ are empty. They coincide with the "bare"
levels and are equally spaced, see the left part of Fig.~\ref{C}. The
 LLs N$\uparrow$ are occupied with filling factors
 $\nu_{0\uparrow}> \nu_{1\uparrow}> \nu_{2\uparrow} >$..., and
$\nu = \nu_{0\uparrow} + \nu_{1\uparrow} + \nu_{2\uparrow} + ...\;$.
Due to the exchange interaction the LLs N$\uparrow$  are shifted down
depending on the level occupation by the energy $\Sigma_{{\rm N}\uparrow}
 = E_{0} \nu_{{\rm N}\uparrow}$, see the right side of Fig.~\ref{C}. If
the excitation is not very strong, and also due to nonparabolicity, one
can consider a three-level model with
$\nu_{0\uparrow} + \nu_{1\uparrow} + \nu_{2\uparrow} = \nu$,
as is shown in Fig.~\ref{C}. Since 
$\nu_{0\uparrow}> \nu_{1\uparrow}> \nu_{2\uparrow}$, the energy shifts 
$\Sigma_{0\uparrow} > \Sigma_{1\uparrow} > \Sigma_{2\uparrow}$, and hence,
the LLs 0$\uparrow$,  1$\uparrow$, and  2$\uparrow$ are nonequidistant. 

The lifetime of the photoelectrons in level  $\:$1$\uparrow$, is
defined by the Auger process
1$\uparrow$ + 1$\uparrow \rightarrow$ 0$\uparrow$ + 2$\uparrow$. 
It is evident that since the LLs are not equidistant this process is
forbidden by energy conservation and can happen only due to the LL
broadening. Hence it is clear that the Auger transitions are well suppressed
if the disbalance in the LLs spacing $\Sigma = (\Sigma_{0\uparrow} - 
\Sigma_{1\uparrow}) - (\Sigma_{1\uparrow} - \Sigma_{2\uparrow})
= E_{0} (\nu - 3 \nu_{1\uparrow})$ is larger than the LL width $\Delta$. 
In this case only the tails of the LLs' density of states (DOS) are
effective. One can suppose that the greatest possibility for energy
conservation occurs when the scattering partners are situated in the 
middle between the centers of the lowest and highest
 LLs, 0$\uparrow$ and 2$\uparrow$.  
At very low excitation, one can assume that in first approximation $\Sigma$
is independent of the excitation intensity. Then, the probability for the
Auger process
1$\uparrow$ + 1$\uparrow \rightarrow$ 0$\uparrow$ + 2$\uparrow$, is
$\tau_{ee}^{-1} \sim \nu_{1\uparrow}$. When $\nu_{1\uparrow}$ is further 
increased the exchange energy $\Sigma_{0\uparrow}$ decreases, while
$\Sigma_{1\uparrow}$  and $\Sigma_{2\uparrow}$ increase, and hence the 
disbalance $\Sigma$ decreases. This obviously causes an increase in
 the scattering rate, resulting in a superlinear 
dependence of $\tau_{ee}^{-1}$ on $\nu_{1\uparrow}$. One can expect that
the nonlinear enhancement of the scattering rate will be essential
when the disbalance change, $\delta \Sigma = 3 E_{0} \nu_{1\uparrow}$,
approaches the LL width $\Delta$, and hence, the crossover
 filling factor is $\nu_{1\uparrow}^{\ast} \simeq \Delta/3 E_{0}$.
Since the LL width, $\Delta \alt 1$ meV\cite{Kuk}, is appreciably
smaller than the exchange energy $E_{0}$, one finds 
$\nu_{1\uparrow}^{\ast} \ll 1$. Thus, the nonlinear dependence of the
scattering rate $\tau_{ee}^{-1}$ on $\nu_{1\uparrow}$ can be pronounced
already at low concentrations of the excited electrons.

As an illustration we calculate the scattering rate $\tau_{ee}^{-1}$ of the
Auger process 
1$\uparrow$ + 1$\uparrow \rightarrow$ 0$\uparrow$ + 2$\uparrow$
using the approaches given in Refs.\cite{Lev,Ts}. We consider a 2DEG in a 
strong magnetic field and a random statistically homogeneous potential with
a correlator: $\langle U(y)U(0) \rangle  =  \Delta^{2}\,
\exp{\bigl(- y^{2}/\Lambda^{2}\bigr)}$, where the correlation length,
$\Lambda$, is much larger than the
magnetic length $l_{B}=(eB/\hbar c)^{1/2}$. The correlation length is of
the order of the spacer $\Lambda \simeq d$, while typical values of the
magnetic length for fields $B$ between 5 and 15 T are $\sim 100\,{\rm \AA}$.
Hence the random potential can be considered as a smooth one in samples with
a spacer $d \geq 200\,{\rm \AA}$. 

We assume that the LLs follow the random potential in space, and the DOS, 
$\varrho(\varepsilon) = (\sqrt{2 \pi} \Delta)^{-1}\exp(-\varepsilon^{2}/2
\Delta^{2})$, where the energy $\varepsilon$ is referenced to the LL 
center, renormalized by the exchange energy. In the calculation
of the scattering rate only relative coordinates of the interacting
electrons are important. We choose the gauge
 $A = (-By,0,0)$. Let $y_{1} = l_{B}^{2} k_{1}$, and
$y_{2} = l_{B}^{2} k_{2}$ be the guiding centers before scattering, and
$y_{1}^{'} = l_{B}^{2} k_{1}^{'}$, $y_{2}^{'} = l_{B}^{2} k_{2}^{'}$ -
the guiding centers after scattering. $k_{1}, k_{2}, k_{1}^{'}, k_{2}^{'}$
are the corresponding momenta. The shifts of the electrons in the scattering 
event are  $(y_{1}^{'} - y_{1}) =  q$, and
$(y_{2}^{'} - y_{2}) = - q$, and the "average" distance between the scattering
partners is $[(y_{2}^{'} + y_{2})/2 - (y_{1}^{'} + y_{1})/2] = p$.
These quantities define the scattering probability. 
The averaged scattering rate of a test electron in the 1$\uparrow$ LL
with an energy $\varepsilon$ referenced to its center is
\begin{eqnarray}
\Bigl\langle \frac{1}{\tau_{ee}} \Bigr\rangle _{\varepsilon} & = &
\int \int_{-\infty}^{+\infty} \frac{dp dq}{2 \pi \hbar \, l_{B}^{2}}
\:| M\,(p,q) - \bar{M}\,(p,q)|^{2}
\Bigl(S_{\varepsilon}(p,q) + 
S_{\varepsilon}(q, p)\Bigr),
\label{R}
\end{eqnarray}
where $M\,(p,q)$ and $\bar{M}\,(p,q)$ are the scattering matrix elements
for the direct and the exchange electrons' collisions, respectively. The
functions $S_{\varepsilon}(p,q)$ and $S_{\varepsilon}(q,p)$ are due to
the statistical factors and the energy conservation. $S_{\varepsilon}(p,q)$
is related to the "deexciting" Auger transition, in which the test
electron is deexcited to the lower level 0$\uparrow$, and its partner is
excited to the upper level 2$\uparrow$, while $S_{\varepsilon}(q,p)$ 
corresponds to the "exciting" transition, in which the test electron is
excited to the upper level 2$\uparrow$, and its partner is deexcited to the
lower level 0$\uparrow$.

It is easy to check that  $\bar{M}(p,q) = M(q, p)$, and  
\begin{eqnarray}
M(p,q) & = & \frac{1}{l_{B}^{2}}\: \int_{-\infty}^{+\infty}
 d\eta \: V( q, \, \eta ) \:  F(q, \, p - \eta),
\label{M} 
\end{eqnarray}
where
\begin{eqnarray}
V(q, \eta) =  \int_{-\infty}^{+\infty} d\xi \: V(\sqrt{\xi^{2} + \eta^{2}})\:
\exp{\Bigl(\frac{iq\xi}{l_{B}^{2}}\Bigr)}
\label{V} 
\end{eqnarray}
is the Fourier transform of the $ee$-interaction potential $V(r)$. 
The $ee$-interaction is chosen as
$V(\sqrt{x^{2} + y^{2}}) = V_{0} \exp\{-(x^{2} + y^{2})/l_{sc}^{2}\}$, where 
$l_{sc}$ is the screening length, $V_{0} \simeq e^{2}/\kappa l_{B}$, and
$\kappa$ is the dielectric constant. 
The function $F(q, p)$ is defined as follows:
\begin{eqnarray}
F(q,\, p) & = & \frac{1}{4}\exp{\Bigl(- \frac{p^{2}}{2 l_{B}^{2}} -
\frac{q^{2}}{2 l_{B}^{2}}
\Bigr)}\, \int_{-\infty}^{+\infty} dz 
e^{-2z^{2}} H_{1}(z+s) H_{1}(z-s) H_{2}(z-r), 
\label{F} 
\end{eqnarray}
where $\:s = (p + q)\,/\,2 l_{B}\:$, $\: r = (p - q)\,/2 l_{B}\:$, and 
$H_{n}(y)$ is the Hermite polynomial. Since the initial and final states
must overlap, the electron shift in the scattering event is of the order of 
or smaller than the magnetic length $l_{B}$, and hence $q \alt l_{B}$. 
 Thus, the Auger transitions are quasivertical in space. 
The "average" distance between interacting electrons is limited by 
the screening length $l_{sc}$ of the $ee$-interaction, $p \alt l_{sc}$. In 
the situation we consider $l_{sc}$ is defined by the electrons in
the 0$\uparrow$ LL, and since this level is only partially occupied,  
the screening is strong and $l_{sc} \simeq l_{B}$\cite{E}. In this case,
the matrix elements $M(p,q)$ and $\bar{M}(p,q)$ that enter Eq.(\ref{R})
are exponentially small if $p,q \gg l_{B}$\cite{Ts}, and therefore the
main contribution to the integrals in Eq.(\ref{R}) arises from 
 $p,q \alt l_{B}$. Note also, that due to the spatial homogeneity,
$M(p, q) = M(-p, -q)$ and $S_{\varepsilon}(p, q) = S_{\varepsilon}(-p, -q)$,
and one can restrict the integration over $p$ in Eq.(\ref{R}) to
 $p > 0$. 

We will consider the low-excitation limit, such that
$\nu_{1\uparrow} \ll \nu_{0\uparrow}$, and $\nu_{2\uparrow} = 0$. In 
this case the electron concentration in the 0$\uparrow$ LL changes only
slightly with pumping, and one can assume that the initial Fermi
distribution in this level is not perturbed. The energy distribution of the
photoelectrons in level 1$\uparrow$ depends on the relation between the 
inter- and intra-LL relaxation times. In order to simplify the calculations
 we will consider the case in which the inter-LL relaxation 
is faster, and thus the excited electrons are not at equilibrium.  
We suppose that they are distributed uniformly in space and are excited in a
rather wide spectral interval, thus their occupation numbers are assumed to be
constant and equal to $\nu_{1\uparrow}$.  

With this in mind the function $S_{\varepsilon}(p,q)$ in Eq.(\ref{R}) is  
\begin{eqnarray}
S_{\varepsilon}(p, q)  = \nu_{1\uparrow} \Bigl\langle
 \delta \Bigl(\varepsilon  + \Sigma + U(p + q) - U(q) - U(p)\Bigr) \:
\Bigl[1 -    f\bigl(U(q)\bigr)\Bigr] \Bigr\rangle,
\label{S}
\end{eqnarray}
where  $\:\langle .... \rangle\:$ stands for the statistical
average\cite{Lev}, and  $f(\epsilon)$  is the Fermi distribution in LL
0$\uparrow$. 
Performing the average over the random potential realizations\cite{Ts}
in Eq.(\ref{S}) at the limit of zero temperature, $T=0$, one obtains
\begin{eqnarray}
S_{ \varepsilon}(p,q)  & = & \frac{\nu_{1\uparrow}}{8\,\sqrt{\pi}\,\Delta} \:
 \frac{\Lambda^{2}}{|pq|} 
\exp\Bigr\{-\frac{1}{4 \Delta^{2}}
\Bigl[\varepsilon - \frac{\Lambda^{2} \Sigma}{2 p q}\Bigr]^{2}\Bigr\}
\:\Bigl\{1 - \Phi\Bigl[-\frac{\Lambda}{2\,\Delta \, |q|}\,
\Bigl(\varepsilon - 
\varepsilon_{F} - \frac{\Sigma q} {2 p}\Bigr)\Bigr]\Bigr\},
\label{S1}
\end{eqnarray}
where $\Phi(x) = \bigl( 2/\sqrt{\pi}\bigr)\;\int_{0}^{x}\,e^{-t^{2}}\,dt\:$  
is the probability integral\cite{GR}, and $\varepsilon_{F}$ is the Fermi
energy referenced to the center of the 0$\uparrow$ LL.

The factor $\{1 - \Phi\}$ in Eq.(\ref{S1}) is due to the occupation of
the 0$\uparrow$ LL. Let us introduce the Fermi-level replica (FLR), which
is given by $\varepsilon_{F} + \hbar \omega_{B} - \Sigma_{1\uparrow}$ and
is thus pinned to level 1$\uparrow$, see Fig.~\ref{N}. The energy 
difference $\varepsilon - \varepsilon_{F}$ in $\Phi$ is the test electron
energy referenced to the FLR. Consider first the Auger transitions with
$q > 0$, i.e., when the scattering partners are closer in space after 
scattering. The factor $\{1 - \Phi\}$ shows that these transitions are strong
if the test electron is above the FLR by an energy 
$\simeq \Sigma/2$, i.e., at $\varepsilon - \varepsilon_{F} \agt \Sigma/2$, and
weak if $\varepsilon - \varepsilon_{F} \alt \Sigma/2$, see the left part
of the Fig.~\ref{N}. Similarly, 
the Auger transitions with $q < 0$ (i.e., when the scattering partners are
closer in space before scattering) are strong at $\varepsilon - \varepsilon_{F}
\agt - \Sigma/2$, and weak if $\varepsilon - \varepsilon_{F} \alt - \Sigma/2$,
see the right part of the Fig.~\ref{N}. In both cases the crossover scale is
$\Delta (l_{B}/\Lambda) \equiv \Delta_{B} \ll \Delta$, much smaller
than the LL width.   
Thus, due to the occupation of the lowest LL,  0$\uparrow$, all Auger
processes for large negative energies of the test electron are quenched.

The origin of the exponential factor is as follows. During the scattering 
event the total momentum and energy are conserved, i.e., 
$[\varepsilon + U(p + q) - 2 \,\Sigma_{1\uparrow}] - 
[U(p) + U(q) - \Sigma_{0\uparrow}] = 0$. Since the Auger transitions are
quasivertical ($q \ll \Lambda $), the random potential $U(y)$ can be
expanded in  powers of $q$ and  $p$. Assuming the test electron is at
$y = 0$, i.e., $\varepsilon = U(0)$, one obtains $U^{''}(0) pq = - \Sigma$,
where $\Sigma  = \Sigma_{0\uparrow} - 2 \:\Sigma_{1\uparrow}$ is the 
disbalance in the LL spacing. The exponential factor in Eq.(\ref{S1}) is
proportional to the conditional probability Probc$\{U(0)=\varepsilon|
U^{''}(0)=-\Sigma/p q \}$ that if the random potential $U(y)$
at $y = 0$ is $\:U(0) = \varepsilon\:$, then its second derivative at the
same point is $\:U^{''}(0) = - \Sigma/p q\:$. Typical magnitudes
of the second derivative of the random potential at energies $|\varepsilon|
\alt \Delta$ are $|U^{''}| \simeq \Delta/\Lambda^{2}$, which is much smaller
than $\Sigma/l_{B}^{2}$ for typical $\Sigma$. Hence energy conservation can
not be satisfied for  $|\varepsilon| \alt \Delta$, and can be obeyed only
in the tails of the DOS where the probability to find large $|U^{''}|$ is
not small. This can be seen from the exponential factor entering
Eq.(\ref{S1}), which has its maximum values at $\varepsilon = \pm 
\varepsilon_{s} \simeq (\Lambda/l_{B})^{2} \Sigma \gg \Delta$, i.e.,
 in the tails of the DOS.

Note, that the results obtained in the case of a smooth random potential
differ from the general predictions given above. Namelly, the scattering rate 
in this case is very sensitive to nonequal distances between the LLs, and
is suppressed not only when the disbalance in the LL spacing is comparable to 
the LL width, but also at much
smaller $\Sigma \agt \Delta(l_{B}/\Lambda)^{2} \equiv \Delta_{s}$ with
$\Delta_{s} \ll \Delta$. 
In addition when the occupation of the excited LL 1$\uparrow$ increases,
the scattering rate responds to a much smaller change of the disbalance than
predicted, $\delta \Sigma \simeq \Delta$.   
Indeed, as follows from the exponential factor in Eq.(\ref{S1}), the scattering
rate $\tau_{ee}^{-1}$ is sensitive to the decrease in $\Sigma$, when
$\delta \Sigma \simeq 8 \Delta_{s}^{2}/\Sigma \ll \Delta$. The 
crossover filling factor is also much smaller,
$\nu_{1\uparrow}^{\ast} \simeq \Delta_{s}^{2}/
\Sigma E_{0} \ll \Delta/E_{0}$. For example, at $\Delta_{s} \simeq 0.1 \;$meV,
$E_{0} \simeq 2 \;$meV, and
$\Sigma \simeq 0.5 \;$meV, $\nu_{1\uparrow}^{\ast}\simeq 0.01$.

The dependence of the scattering rate on the concentration of the excited
electrons at the test electron energy  $\varepsilon = \varepsilon_{F}$
is shown in Fig.~\ref{T}. The magnetic field is  $B$ = 6 T, and
the curves correspond to two different electron concentrations:
 $n=4.5 \times 10^{10}$ cm$^{-2}$, i.e., $\nu = 0.31$ (curve 1), and 
 $n=7.3 \times 10^{10}$ cm$^{-2}$, i.e., $\nu = 0.5$ (curve 2). The 
other parameters are as follows: $\kappa = 12$,
$\Delta = 1$ meV, $\Lambda = 300\,{\rm \AA}$, and $E_{0} = 2$ meV\cite{Dp}.
From Fig.~\ref{T} it can be seen that in both cases the scattering 
rate $\tau_{ee}^{-1}$ changes linearly with the filling factor 
$\nu_{1\uparrow}$ only at very small $\nu_{1\uparrow}$. The arrows in
Fig.~\ref{T} indicate the crossover values of $\nu_{1\uparrow}^{\ast}$, when
the deviation from the linear law accounts for about $100 \%$.
For a given $\nu_{1\uparrow}$, the scattering rate $\tau_{ee}^{-1}$ is
more suppressed at larger electron concentration, because of the larger 
disbalance in the LL spacing. 

We are thankful to E.Gornik for fruitful discussions. This work is supported
by the MINERVA Foundation.

\newpage
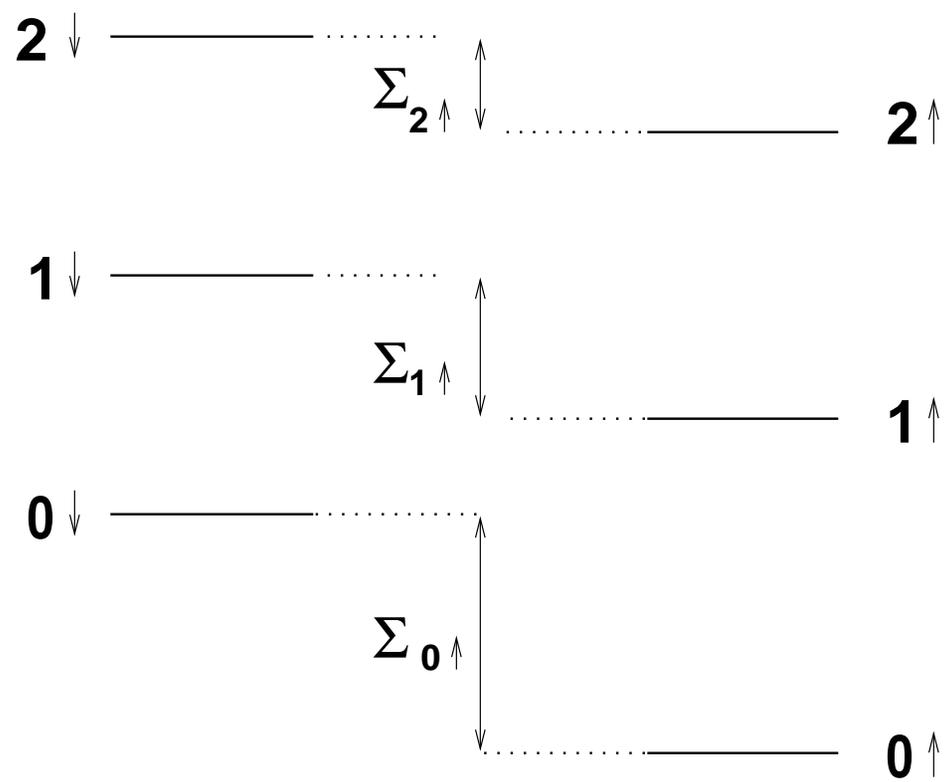
\begin{figure} 
\caption{The Landau level ladder in the spin polarized case.} 
\label{C} 
\end{figure}
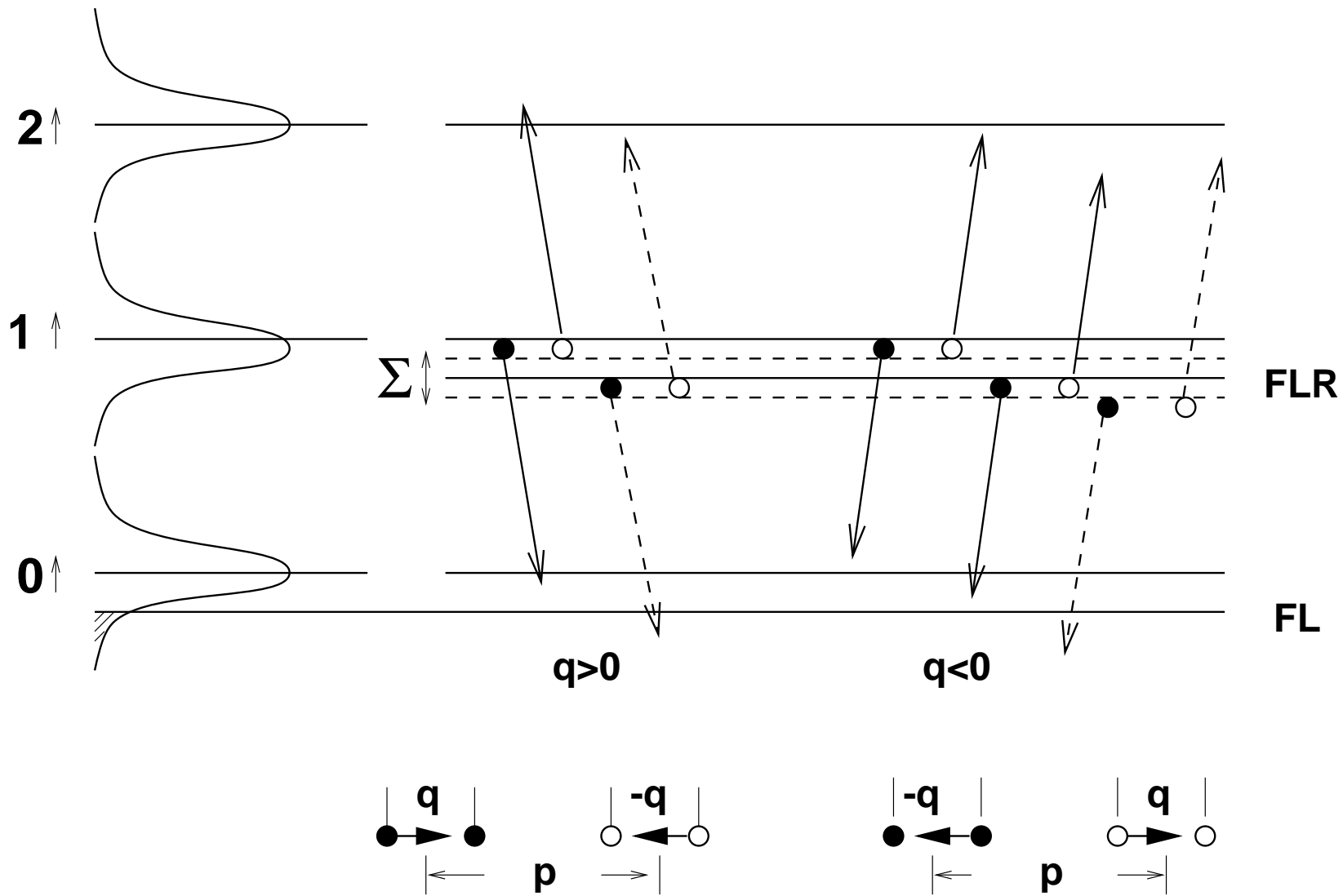
\begin{figure}
\caption{The Auger process:
1$\uparrow$ + 1$\uparrow \rightarrow$ 0$\uparrow$ + 2$\uparrow$. The
suppresed processes are shown with dashed arrows.} 
\label{N} 
\end{figure}
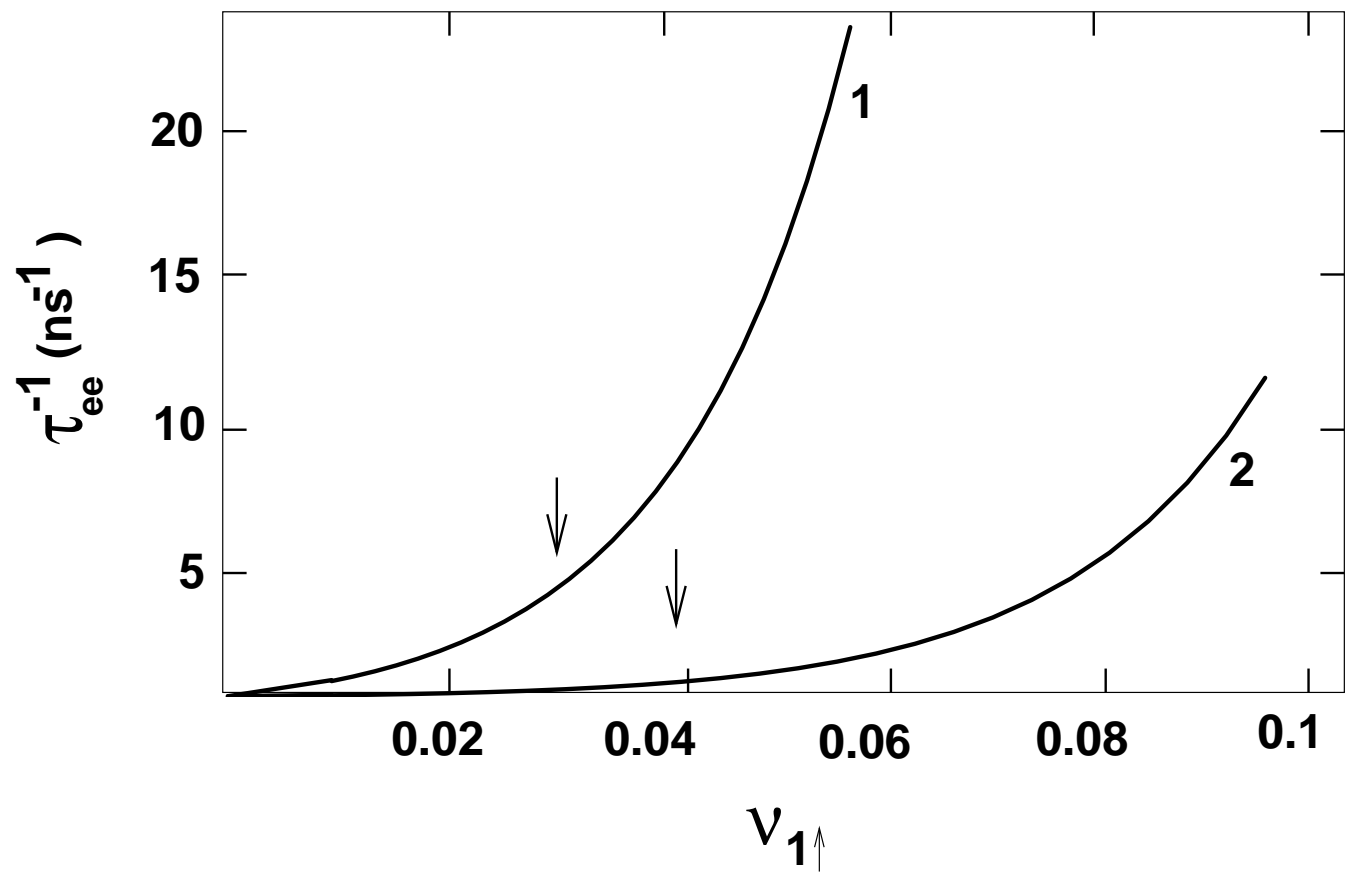
\begin{figure} 
\caption{The Auger scattering rate, $\tau_{ee}^{-1}$,  for the process:
1$\uparrow$ + 1$\uparrow \rightarrow$ 0$\uparrow$ + 2$\uparrow$ at
$\varepsilon = \varepsilon_{F}$,  $\nu = 0.31$ (curve 1), and
$\nu = 0.5$ (curve 2), versus the filling factor $\nu_{1\uparrow}$.  The
arrows indicate the crossover filling factors $\nu_{1\uparrow}^{\ast}$
(see also text).}
\label{T} 
\end{figure}
\end{document}